\documentclass[lettersize,journal]{IEEEtran}
\pdfoutput=1
\usepackage{amsmath,amsfonts}
\usepackage{algorithmic}
\usepackage{algorithm}
\usepackage[english]{babel}
\usepackage{amsthm}
\usepackage{array}
\usepackage[caption=false,font=normalsize,labelfont=sf,textfont=sf]{subfig}
\usepackage{textcomp}
\usepackage{color,soul}
\usepackage{url}
\usepackage{verbatim}
\usepackage[utf8]{inputenc}
\usepackage{amsmath,bm}
\usepackage{graphicx}
\usepackage{bbding}
\usepackage{pifont}
\usepackage{wasysym}
\usepackage{placeins}
\usepackage{cite}
\usepackage[short]{optidef}
\usepackage{mathrsfs}
\usepackage{xcolor}
\usepackage{hyperref}
\usepackage{multirow}

\usepackage{cite}
\usepackage{amsmath,amssymb,amsfonts}
\usepackage{algorithmic}
\usepackage{graphicx}
\usepackage{textcomp}
\usepackage{multirow}
\usepackage{amsmath}
\usepackage{booktabs}

\hypersetup{backref=true,
	pagebackref=true,
	hyperindex=true,
	colorlinks=true,
	breaklinks=true,
	urlcolor= black,
	linkcolor= black,
	bookmarks=true,
	bookmarksopen=false,
	filecolor=black,
	citecolor=black,
}

\theoremstyle{lemma}

\begin{document}
	
	\title{
	Optimized Deep Feature Selection for Pneumonia Detection: A Novel RegNet and XOR-Based PSO Approach}

\author{
	Fatemehsadat Ghanadi Ladani, Samaneh Hosseini Semnani

	\thanks{
		F. G. Ladani and S. H. Semnani are with the Department of Electrical and Computer Engineering, Isfahan University of Technology, Isfahan, (e-mails: f.ghanadi@ec.iut.ac.ir, samaneh.hoseini@iut.ac.ir).

	}
	\thanks{
}}



\maketitle

\begin{abstract}

	Pneumonia remains a significant cause of child mortality, particularly in developing countries where resources and expertise are limited. The automated detection of Pneumonia can greatly assist in addressing this challenge. In this research, an XOR based Particle Swarm Optimization (PSO) is proposed to select deep features from the second last layer of a RegNet model, aiming to improve the accuracy of the CNN model on Pneumonia detection. The proposed XOR PSO algorithm offers simplicity by incorporating just one hyperparameter for initialization, and each iteration requires minimal computation time. Moreover, it achieves a balance between exploration and exploitation, leading to convergence on a suitable solution. By extracting $163$ features, an impressive accuracy level of $98\%$ was attained which demonstrates comparable accuracy to previous PSO-based methods. The source code of the proposed method is available in the GitHub repository\footnote{https://github.com/fatemehghanadi/XORPSO}.
	
\end{abstract}


\section{Introduction}


\IEEEPARstart{P}{neumonia} is a significant health challenge, especially in developing countries. Each year, around $1.5$ million people suffer from pneumonia, with $450,000$ losing their lives tragically. Additionally, an astonishing $10.5\$$ billion is spent on hospital care for these individuals \cite{makhnevich2019clinical}. Pneumonia can lead to both quick death and prolonged health problems. It primarily affects vulnerable groups like children and older adults with existing chronic conditions. The causes of pneumonia include various microorganisms such as bacteria, viruses, and fungi. People with pneumonia often experience respiratory and systemic symptoms, which require a diagnostic approach considering both clinical signs and radiological findings. Timely identification and proper use of antimicrobial treatments are crucial, as delays can result in unfavorable outcomes \cite{Torres2021}.

Chest X-ray (CXR) imaging stands as the primary method for screening, prioritizing, and diagnosing diverse forms of pneumonia. In the midst of flu season, viral pneumonia takes center stage, and CXR imaging emerges as a pivotal asset in immediate patient management. Radiologists possess insights into distinct CXR characteristics that may signal the likelihood of viral pneumonia diagnosis \cite{wang2021deep}.

Over the past few years, artificial intelligence (AI) has introduced fresh approaches that significantly speed up the development of radiological diagnosis tools. When it comes to identifying common lung diseases from chest X-rays (CXRs), AI techniques like weakly supervised classification or attention-focused convolutional neural networks are being used \cite{wang2021deep}.

This paper introduces an XOR PSO approach for binary feature selection to improve the accuracy of Pneumonia detection. The RegNet model, which was pretrained on the ImageNet dataset, is utilized as the CNN architecture. By fine-tuning the RegNet model on a Pneumonia dataset, the XOR PSO effectively selects highly informative features from the second last layer. The proposed method demonstrates superior performance in terms of execution time and simplicity compared to other PSO-based algorithms.
The rest of the paper is organized as follows:
Section $2$ reviews the main methods proposed in the domain of
Pneumonia and COVID-19 detection from CXRs. Section $3$ gives a detailed description of the proposed method, Section $4$ reports the experiments and the corresponding analysis, and Section $5$ is the conclusion.

\section{Related work}
Methods that performed well in accurately detecting pneumonia often showed similar success in detecting COVID-19 when trained on COVID-19 dataset. Therefore, in this section, we provide a review of CNNs and feature selection methods used for detecting both COVID-19 and Pneumonia. CNNs are used to find the informative patterns and feature selection methods are employed for extracting the best features from CNN output layer.

\subsection{CNNs for CXRs}

Huang et al. \cite{huang2022lightweight} conducted a study to evaluate the performance of various convolutional neural network (CNN) models on chest X-ray image datasets for COVID-19 detection. The results showed that the InceptionV3 model achieved the highest accuracy of $96.50\%$ before fine-tuning, while the EfficientNetV2 model surpassed other models after fine-tuning, with an accuracy of $97.73\%$. Notably, Huang et al. proposed a novel model called LightEfficientNetV2, which demonstrated accuracy of $98.33\%$ on chest X-ray images. These findings highlight the superior performance of the LightEfficientNetV2 model in accurately identifying COVID-19 cases, surpassing the results achieved by previous state-of-the-art models.

Gayathri J.L. et al. \cite{gayathri2022computer} developed a CAD model for COVID-19 detection using chest X-ray images. The study utilized multiple pre-trained networks in combination with a sparse autoencoder for dimensionality reduction. The model employed a Feed Forward Neural Network (FFNN) for COVID-19 detection. The training dataset consisted of 504 COVID-19 images and $542$ non-COVID-19 images. The proposed model achieved an accuracy of $95.78\%$ and an AUC of $98.21\%$ when the InceptionResnetV2 and Xception networks were combined. The study also demonstrated that incorporating sparse autoencoder as a dimensionality reduction technique improved the accuracy of the model.

\subsection{Feature Selection Methods}
Pramanik et al. \cite{pramanik2022adaptive} implemented a feature selection technique in their Computer-Aided Detection (CAD) system for Pneumonia detection from Chest X-rays. The feature selection process involved utilizing the pre-trained ResNet50 model to extract deep features and could achieve accuracy of $96.49\%$. However, not all of these features were informative for Pneumonia detection. To address this, they employed a modified version of PSO called Adaptive and Altruistic PSO (AAPSO). The AAPSO algorithm was designed to select the most relevant and informative features from the ResNet50 model. It incorporated a memory-based adaptation parameter and introduced an altruistic behavior between particles. This approach allowed the AAPSO algorithm to effectively eliminate non-informative features obtained from the ResNet50 model. By applying the AAPSO feature selection method, this method enhanced the Pneumonia detection ability of their CAD system and could achieve accuracy of $98.41\%$. The selected features captured the important patterns and characteristics related to Pneumonia detection, thereby improving the overall performance and accuracy of the CAD system in identifying cases of Pneumonia from Chest X-ray images.

In another study Narin \cite{narin2021accurate} focuses on developing an automatic diagnosis and detection system for COVID-19 and Pneumonia using X-ray images. The aim is to provide an additional tool to control the epidemic and support the existing RT-PCR test. Three different CNN models (ResNet50, ResNet101, InceptionResNetV2) were trained on X-ray images from three classes: COVID-19, Normal, and Pneumonia. The researchers compared the performance of two feature selection methods: PSO algorithm and ant colony algorithm (ACO). Support vector machines (SVM) and k-nearest neighbor (k-NN) classifiers were used with a 10-fold cross-validation method to evaluate the results. The highest overall accuracy achieved was $99.83\%$ with the SVM algorithm without feature selection. However, the highest performance of $99.86\%$ was obtained after applying the feature selection process using the SVM and PSO method. The study demonstrates that incorporating feature selection not only enhances the system's performance but also reduces the computational load by utilizing less features for classification. By providing higher accuracy in COVID-19 detection, this system can contribute to efficient diagnosis and control of the epidemic.

Canayaz \cite{canayaz2021mh} aimed to develop a deep learning-based approach for the early diagnosis of COVID-19, which is crucial for saving lives. A dataset consisting of three classes (COVID-19, Normal, and Pneumonia) was created. After applying image pre-processing methods, deep learning models, including AlexNet, VGG19, GoogleNet, and ResNet, were utilized for feature extraction from the enhanced dataset. To select the most effective features, two metaheuristic algorithms (Binary PSO and binary gray wolf optimization) were employed. The selected features were then classified using SVM. The proposed approach achieved an impressive overall accuracy of $99.38\%$. The results demonstrated that the combination of deep learning models, feature selection using metaheuristic algorithms, and SVM classification can significantly contribute to COVID-19 diagnostic studies. The high accuracy obtained through the proposed approach suggests its potential as a valuable tool to assist experts in diagnosing COVID-19 at an early stage.

\section{Proposed Method}
\begin{figure}
	\centering
	\includegraphics[width=0.5\textwidth]{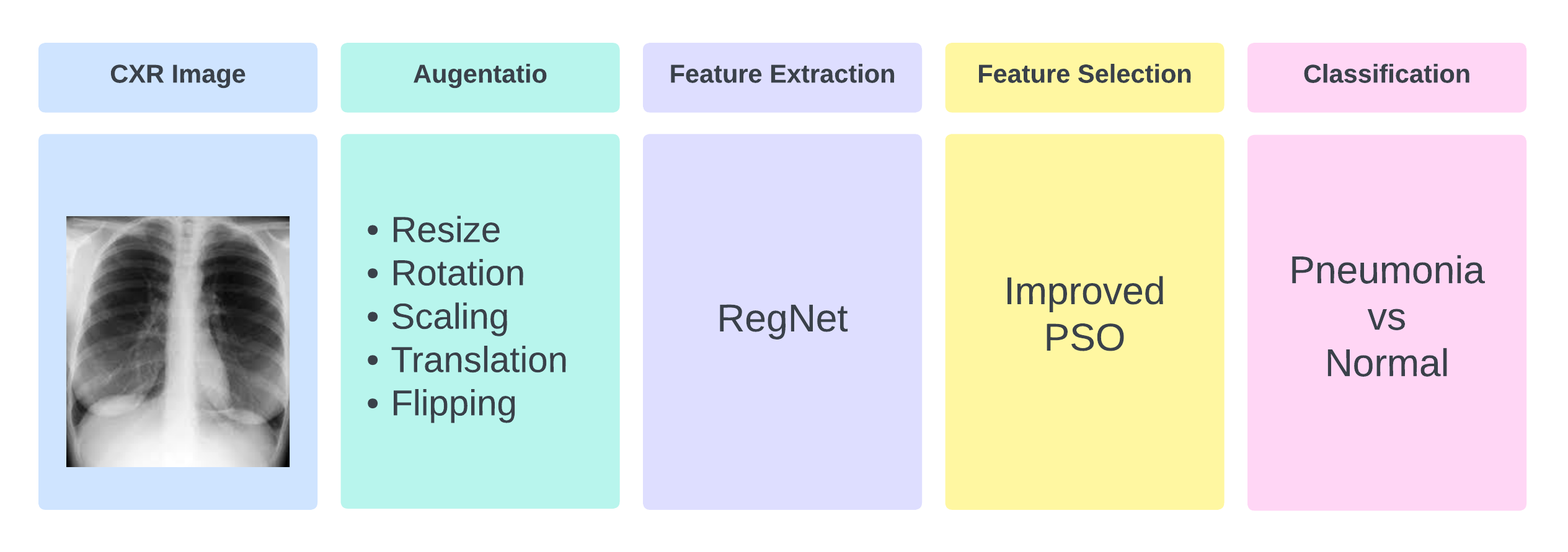}
	\caption{The overal pipeline of the proposed method.}		\label{fig1}
\end{figure}

This section presents a comprehensive overview of the implemented pipeline in this research, which is visualized in Figure \ref{fig1}. The pipeline encompasses several key steps. Initially, the input images are resized to dimensions of $(224, 224)$ pixels, and the augmentation techniques such as rotation, horizontal and vertical flip, scaling and translation are applied on them. These augmentation methods generate a larger set of images, enhancing the model's ability to learn features effectively. Subsequently, this augmented image set is utilized as input for the chosen CNN model, namely RegNet, which could achieve less error rate than ResNet50, ResNet101,  DenseNet-169 and SE-ResNet50 on CIFAR-10 dataset \cite{xu2022regnet}. This model has been pre-trained on the ImageNet dataset. To further enhance the model's performance, a fully connected layer consists of $512$ neurons is incorporated. The output layer, comprising two neurons, is responsible for determining the probability of each class. For enhanced accuracy, the input for feature selection is derived from the utilization of the $512$ neurons from the second last layer. For feature selection, the XOR PSO algorithm is employed. This algorithm is specifically tailored for the task of feature selection, treating it as a discrete binary feature selection problem. The XOR PSO algorithm is enable to efficiently identify and select the most relevant features which improve the classification accuracy and decrease computational burden. The k-NN classifier utilizes the selected features to classify and make predictions accurately. In this section the main components of the proposed method are discussed. It includes dataset, RegNet as desired CNN, PSO as inspired algorithm for continuous optimization, XOR PSO as proposed optimizer for discrete binary feature selection, Mutual Information as a method for initializing XOR PSO and fitness function for validating performance of the XOR PSO.

\subsection{Dataset}
In this research, the publicly available CXR Pneumonia dataset\footnote{https://www.kaggle.com/datasets/paultimothymooney/chest-xray-pneumonia} is utilized for training the CNN models \cite{kermany2018identifying}. This dataset, however, presents a challenge of class imbalance, wherein the number of images in each class varies significantly. The distribution of images in each class is summarized in Table \ref{table1}, providing an overview of the dataset's class distribution.

\begin{table}[htb]
	\centering
	\caption{\textbf{Class and Sample Distribution}}
	\label{table1}
	\setlength{\tabcolsep}{10pt}
	\begin{tabular}{@{}lllll@{}}
		\toprule
		\textbf{Class} & \textbf{Setting} & \textbf{Samples}   \\ \midrule
		\textbf{Normal} & Train & 1260  \\ 
		& Test & 324 \\ 
		\textbf{Pneumonia} & Train & 3317  \\ 
		& Test & 831 \\ 
		\bottomrule
	\end{tabular}
\end{table}

\subsection{RegNet}
ResNet and its related models (such as Inception-ResNet \cite{Szegedy_Ioffe_Vanhoucke_Alemi_2017} and ResNeXt \cite{Xie2016}) have demonstrated remarkable success in various image classification tasks, including the well-known ImageNet dataset. Despite their achievements, these models face limitations in exploring new and potentially valuable features due to their simplistic shortcut connection mechanism. To overcome this limitation, the RegNet models were developed, introducing a regulator module as a memory mechanism to extract complementary features. The regulator module utilizes convolutional RNNs, which excel at extracting rich and informative representations. These RNNs capture spatio-temporal information, enabling them to uncover intricate patterns and relationships within the data. By incorporating the regulator module into the ResNet architecture, RegNet models go beyond the traditional approach and harness the power of complementary feature extraction. This enhancement significantly improves the model's ability to capture diverse and discriminative information, ultimately leading to better accuracy in image classification tasks, including the challenging ImageNet dataset \cite{xu2022regnet}.

\subsection{Particle Swarm Optimization}
PSO, originally introduced by Kennedy and Eberhart \cite{kennedy1995particle}, is an iterative population-based optimization algorithm known for its simplicity, minimal number of controlling parameters, and computational efficiency. In PSO, a population of particles moves through a continuous search space to find the optimal solution to a given problem. Each particle represents a potential solution. The mathematical formulation of PSO is represented by Equations \ref{e6} and \ref{e7}.

\begin{equation}
	V_{t+1} = W \cdot V_{t} + C_{1}R_{1} \cdot (P_{\text{best}, t} - X_{t}) + C_{2}R_{2} \cdot (G_{\text{best}, t} - X_{t}) \label{e6}
\end{equation}

\begin{equation}
	X_{t+1} = X_{t} + V_{t+1}
	\label{e7}
\end{equation}

In Equation \ref{e6}, the terms ${R}_{1}$ and ${R}_{2}$ represent random numbers that are generated within the range of (0,1). ${C}_{1}$(Cognitive Coefficient) and ${C}_{2}$(Social Coefficient) are parameters that control the influence of the individual particle's best-known position (personal best) and the global best-known position (global best), respectively, on the particle's movement and exploration within the search space. ${C}_{1}{R}_{1}.({P}_{best,t}-{X}_{t})$ term corresponds to the self-confidence term, while ${C}_{2}{R}_{2}.({G}_{best,t}-{X}_{t})$ represents the swarm confidence term. The variable ${X}_{t}$ and ${V}_{t}$ denote the position and velocity at time $t$, $W$ is the inertia rate and ${P}_{best,t}$ and ${G}_{best,t}$ denote the personal and global best position at time $t$ in the PSO algorithm. In PSO algorithm, each particle endeavors to navigate towards both its individual best-known point (${P}_{best}$) and the globally best-known point (${G}_{best}$). It refines its ${P}_{best}$ if it uncovers a superior solution, or it refines the ${G}_{best}$ if it discovers a solution superior to those found by other particles.

\subsection{XOR PSO}
The conventional PSO algorithm is designed for searching in continuous spaces to find optimal solutions. However, feature selection problems involve discrete spaces, requiring modifications to adapt PSO for such scenarios. In this context, the concept of velocity and position needs to be redefined. For the sake of feature selection, each position in the PSO algorithm represents a potential solution, depicted as a binary vector that indicates which features should be selected. The velocity component directly influences the movement of positions. For binary feature selection we proposed the velocity to be calculated using the $XOR$ logic function to capture the disparity between the global best position, personal best position, and the current position. If $V$ is equal to $1$, the $X$ will change. Also if ${P}_{best}$ or ${G}_{best}$ is equal to $X$, the $XOR$ $({P}_{best}$ $or$ ${G}_{best},$ $X)$ will be $0$ (Table \ref{table4}, \ref{table5}). By implementing these adjustments, the XOR PSO algorithm adepts at efficiently exploring and optimizing feature selection within discrete spaces.

\begin{table}[!ht]
	\centering
	\caption{Calculating $XOR($${\textbf{P}}_{\textbf{best}}$ or ${\textbf{G}}_{\textbf{best}}$, $\textbf{X})$.}
	\label{table4}
	\begin{tabular}{@{}lllll@{}}
		\toprule
		\textbf{$\textbf{XOR}$} & \textbf{${\textbf{P}}_{\textbf{best}}$ or ${\textbf{G}}_{\textbf{best}} $} & \textbf{X}   \\ \midrule
		\textbf{0} & 0 & 0   \\ 
		\textbf{1} & 0 & 1 \\ 
		\textbf{1} & 1 & 0   \\ 
		\textbf{0} & 1 & 1 \\ \bottomrule
	\end{tabular}
\end{table}

\begin{table}[!ht]
	\centering
	\caption{Using XOR for updating \textbf{X}.}
	\label{table5}
	\begin{tabular}{@{}lllll@{}}
		\toprule
		\textbf{${\textbf{X}}_{\textbf{t+1}}$} & \textbf{${\textbf{X}}_{\textbf{t}}$} & \textbf{${\textbf{V}}_{\textbf{t+1}}$}   \\ \midrule
		\textbf{0} & 0 & 0   \\ 
		\textbf{1} & 0 & 1 \\ 
		\textbf{1} & 1 & 0   \\ 
		\textbf{0} & 1 & 1 \\ \bottomrule
	\end{tabular}
\end{table}

To calculate the velocity in the XOR PSO algorithm, the inertia weigh is employed as the sole control parameter. Proposed method reduces the necessity of adjusting multiple control parameters. Inertia weight ($W$) which is between $0$ and $1$ determines the influence of the previous velocity on the updated velocity. Additionally, a random number (${R}_{1}$), ranging from $-1$ to $1$ is introduced to multiply with the difference between the personal best position and the current position. This random factor allows the particle to randomly move closer to or farther away from its best position. In contrast, the second random number (${R}_{2}$), which multiplies with the difference between the global best position and the current position, remains between $0$ and $1$, following the traditional PSO approach. 
Then the updated velocity (${V}_{t+1}$) is mapped to $0$ if it is less than $0.5$, and to $1$ if it exceeds $0.5$. In conventional PSO $C1$ and $C2$ in self confidence and swarm confidence terms are between $0$ and $1$ and influence the balance between focusing on individual's best solutions and the swarm's best solutions. As the threshold of $0.5$ determines the ${V}_{t+1}$ value to become $0$ or $1$ in XOR PSO, setting $C1$ and $C2$ more than $1$ will increase the probability of ${V}_{t+1}$ to become $1$ and setting them less than $1$ will decrease this probability. So $C1$ and $C2$ terms were omitted or this is like they were set to $1$.

The formulation of XOR PSO algorithm is represented by equations \ref{e2}, \ref{e3} and \ref{e4}.

\begin{equation}
	V_{t+1} = W \cdot V_{t} + R_{1} \cdot \text{XOR}(P_{\text{best}, t}, X_{t}) + R_{2} \cdot \text{XOR}(G_{\text{best}, t}, X_{t}) \label{e2} 
\end{equation}

\begin{equation}
	V_{t+1} = \begin{cases} 
		0 & \text{if } V_{t+1} < 0.5 \\
		1 & \text{if } V_{t+1} \geq 0.5
	\end{cases} \label{e3}
\end{equation}

\begin{equation}
	X_{t+1} = \text{XOR} (X_{t}, V_{t+1}) \label{e4}
\end{equation}

\subsection{Mutual Information}
Mutual information (MI) is utilized to identify the most important features. The extent of the relationship between two variables is measured, and the shared information between them is assessed. By calculating the mutual information between different variables, the features that had a strong influence on the outcome could be determined. The degree of dependence between variables is quantified, and the ones that contributed the most to the classification are identified. With the help of mutual information, the reduction in uncertainty of one variable is assessed when the other variable is known. This information played a crucial role in initializing subset of the population for the PSO algorithm.
The formula for calculating MI between two discrete random variables X and Y is represented by Equation \ref{e1}.

\begin{equation}
	MI(X, Y) = \sum_{y}\sum_{x} P(x, y) \log\left(\frac{P(x, y)}{P(x) \cdot P(y)}\right) \label{e1}
\end{equation}

\subsection{Fitness function}

To evaluate the strength of a candidate solution, a fitness function needs to be defined, which determines the quality of the solution. If accuracy alone is considered as the fitness function, the solution may contain a large number of features, some of which may not be informative, but still achieve the best accuracy. On the other hand, if fitness is considered as a ratio between accuracy and the number of selected features, the convergence speed becomes faster, as the particles focus on reducing the number of features rather than optimizing accuracy. To address this issue, the fitness function is updated in two phases. In disease detection, prioritizing accuracy over the number of remaining features is crucial. Initially, the primary objective is to achieve a high level of accuracy. Once a satisfactory accuracy level is attained, the algorithm shifts its focus towards reducing the number of features while maintaining the previous level of accuracy. In this problem, the accuracy threshold were set to $0.98$. If the current particle accuracy is lower than $0.98$, the fitness is set as the accuracy. However, if the current accuracy is equal to or greater than $0.98$ , the fitness is calculated as Equation \ref{e8}. In this equation the left out feature ratio is subtracted from $2$ to achieve a fitness number more than $1$ and consider particles with accuracy more than $0.98$ as best particles and focus on decreasing the number of their features. The approach ensures that particles strive for more informative features when their  resulted accuracy is below $0.98$, and emphasis on utilizing fewer features when their resulted accuracy exceeds $0.98$. In other words, it promotes the exploration of more informative features for particles which result in lower accuracy and encourages the exploitation to decrease the number of feature for particles already achieving high accuracy. This XOR PSO is shown in Algorithm \ref{Alg1}. The effects of using two different fitness function on the number of selected features is represented in Figure \ref{fig8}.

\begin{equation}
	\text{Fitness}= 2 - \frac{\text{number of selected features}}{\text{total number of features}}
	\label{e8}
\end{equation}

\begin{algorithm}[!t]
	\caption{XOR PSO}
	\label{Alg1}
	\begin{algorithmic}[1]
		\FOR{Iteration in Iterations}
		\FOR{particle in population}
		\STATE Update $V$ using Equations \ref{e2} and \ref{e3}
		\STATE Update $X$ using Equation \ref{e4}
		\STATE $Classifier.fit(TrainData(selected\_features))$
		\STATE $particle.Accuracy  =  Classifier.evaluate(ValData(selected\_features))$
		\IF{$particle.Accuracy < AccThreshold$}
		\STATE $particle.Fitness = particle.Accuracy$
		\ELSE
		\STATE Calculate $particle.Fitness$ using Equation (\ref{e8})
		\ENDIF
		\STATE Update $Gbest$
		\STATE Update $particle.Pbest$
		\STATE Update $Best\_Acc$
		\ENDFOR
		\ENDFOR
		\STATE Return $Gbest$
	\end{algorithmic}
\end{algorithm}

\section{Results and analysis}
This section focuses on the results obtained by applying the proposed method on the Pneumonia dataset. All the experiments have been performed
on Google Colab which is a NVIDIA Tesla K80 with 12GB of VRAM. The performance and outcomes of the approach are thoroughly examined and discussed, shedding light on the findings.

\subsection{Performance of CNN}
In the present study, a comparison is conducted among several pre-trained CNN architectures for the detection of Pneumonia from chest X-ray images. These architectures have been specifically designed for image classification and pattern recognition tasks.

\begin{figure}
	\centering
	\includegraphics[width=0.4\textwidth]{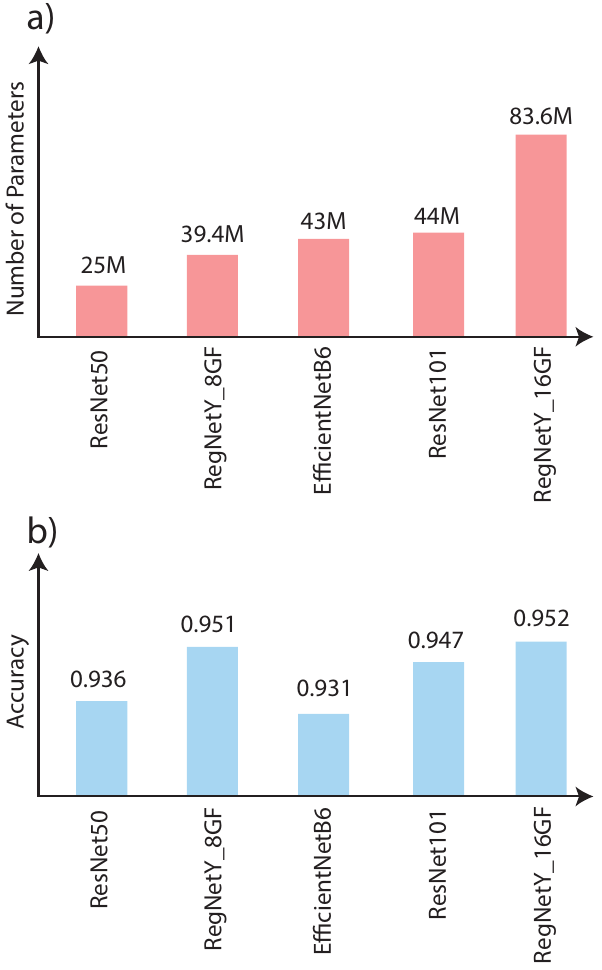}
	\caption{Different CNN models, a) Number of parameters. b) Accuracy.}
	\label{fig2}
\end{figure}

The CNN models included ResNet \cite{he2016deep}, known for their high accuracy on the ImageNet dataset prior to 2015, with variations comprising 50 or 101 layers, EfficientNetB6 \cite{tan2019efficientnet}, a more recent architecture that outperformed ResNet on the ImageNet dataset in 2020, and RegNet \cite{xu2022regnet}, which exhibited superior performance compared to ResNet50 while maintaining a relatively small increase in the number of parameters, were employed. To evaluate the performance of these models, a validation set representing $20\%$ of the total dataset was utilized. The resulting outcomes, illustrated in Figure \ref{fig2}, provide a clear indication of the accuracy achieved by each individual model.

\subsection{Hyperparameter tuning}
Hyperparameters play a crucial role in the training of deep learning models, as they significantly influence the model's learning process. However, determining the optimal values for these hyperparameters can be a challenging task. Among the various hyperparameters, the batch size and learning rate hold particular importance. Selecting an optimal learning rate is critical to strike a balance between fast convergence and avoiding overshooting or getting stuck in suboptimal solutions.
\subsubsection{CNN parameters}
\begin{figure}
	\centering
	\includegraphics[width=0.5\textwidth]{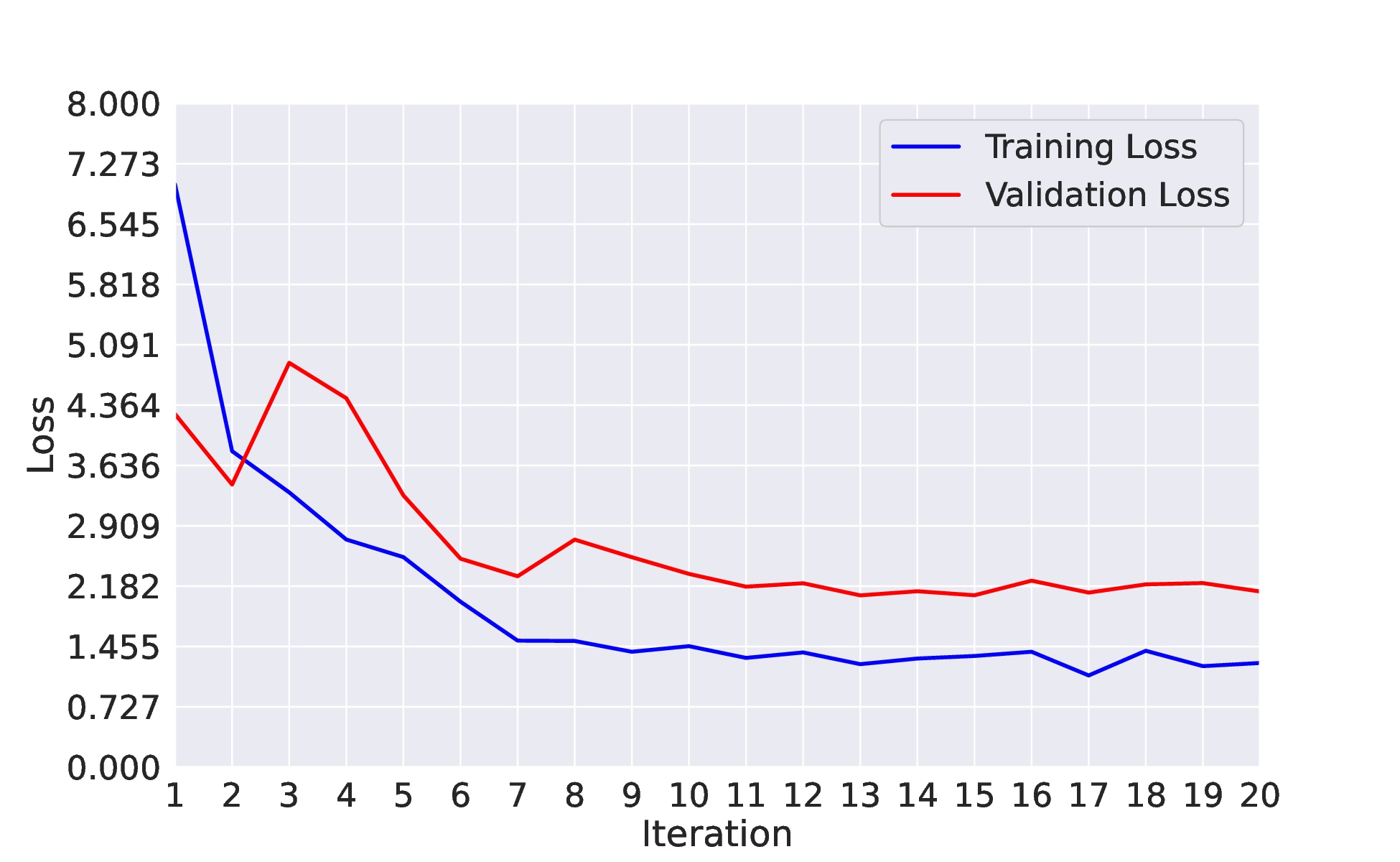}
	\caption{RegNet Loss curves.}
	\label{fig5}
\end{figure}

\begin{figure}
	\centering
	\includegraphics[width=0.5\textwidth]{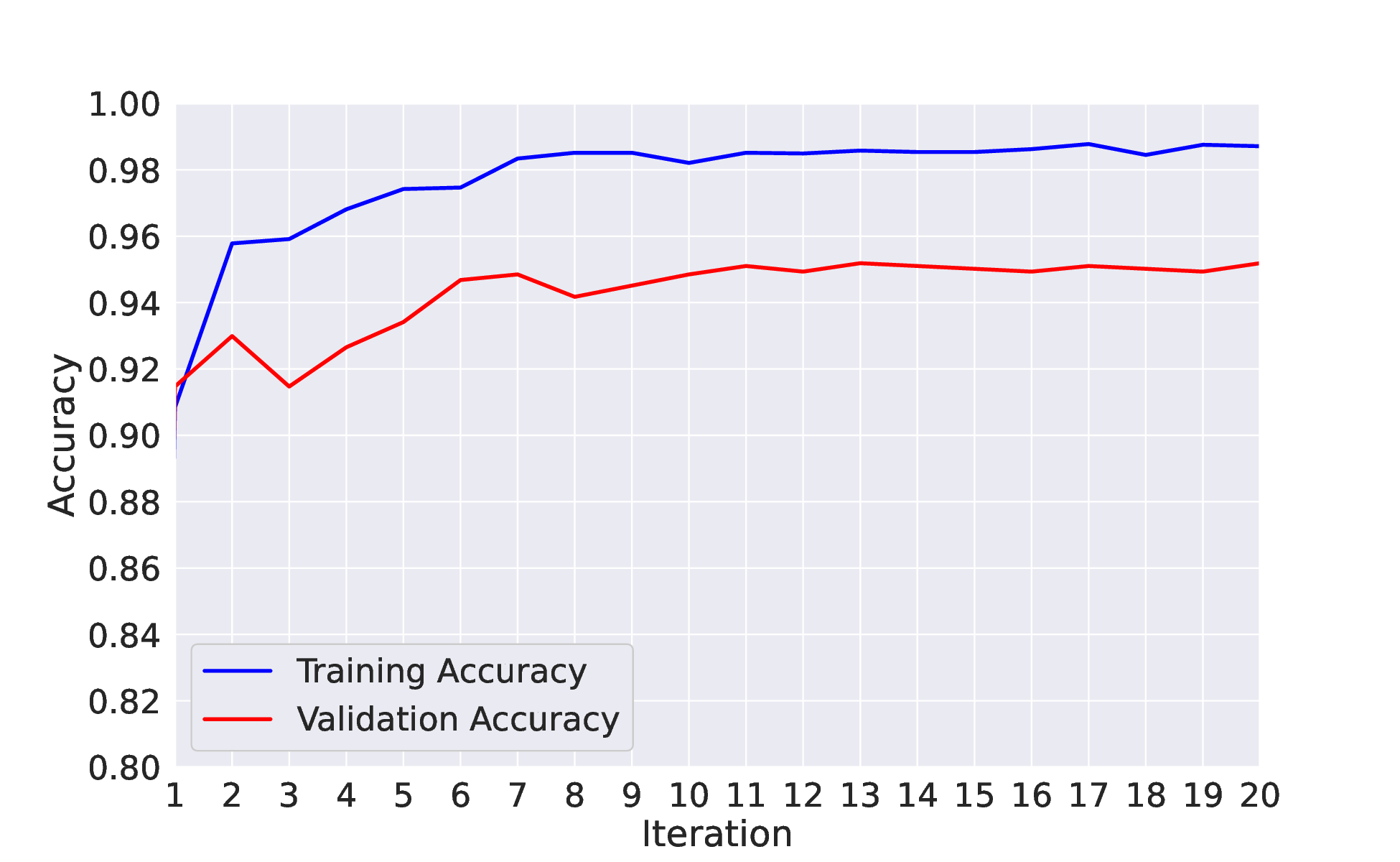}
	\caption{RegNet Accuracy curves.}
	\label{fig6}
\end{figure}

To determine the appropriate values for the learning rate and batch size, the parameters of \cite{pramanik2022adaptive} is utilized. To promote smoother learning and reduce overfitting, the learning rate is set to $0.0001$ and decreased by a factor of ten after the completion of every five epochs and batch size set as $32$. For the optimization of the deep learning model, the Adam optimizer is utilized. This optimizer is widely recognized and performs well in various tasks. Additionally, the cross-entropy loss function, which is commonly used in classification problems, is employed to measure the model's performance. The results are shown in Figures \ref{fig5} and \ref{fig6}.

\subsubsection{Feature selection}
The XOR PSO algorithm is employed to enhance the accuracy of the classification model. One of the advantages of the proposed XOR PSO algorithm is its simplicity, as it requires only the adjustment of the inertia weight. In order to improve the exploitation phase of the algorithm, the inertia weight is initially set to $1$ and it gradually decrease by $0.05$ every $5$ iterations. This approach allows for more exploration in the early iterations and greater focus on exploitation in the later iterations.

\begin{figure}
	\centering
	\includegraphics[width=0.4\textwidth]{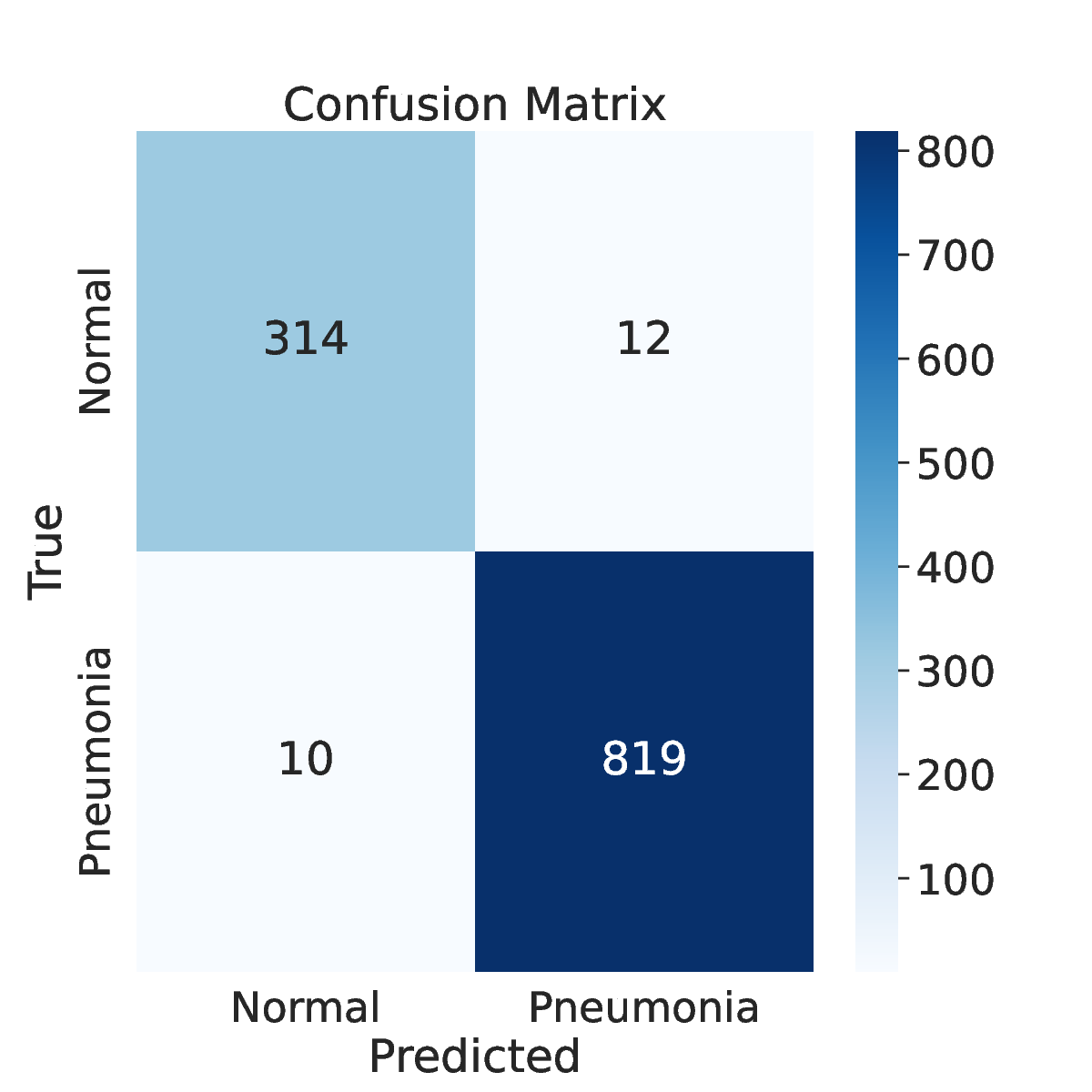}
	\caption{Confusion Matrix for Results of the Proposed Method.}
	\label{fig3}
\end{figure}

\begin{figure}
	\centering
	\includegraphics[width=0.5\textwidth]{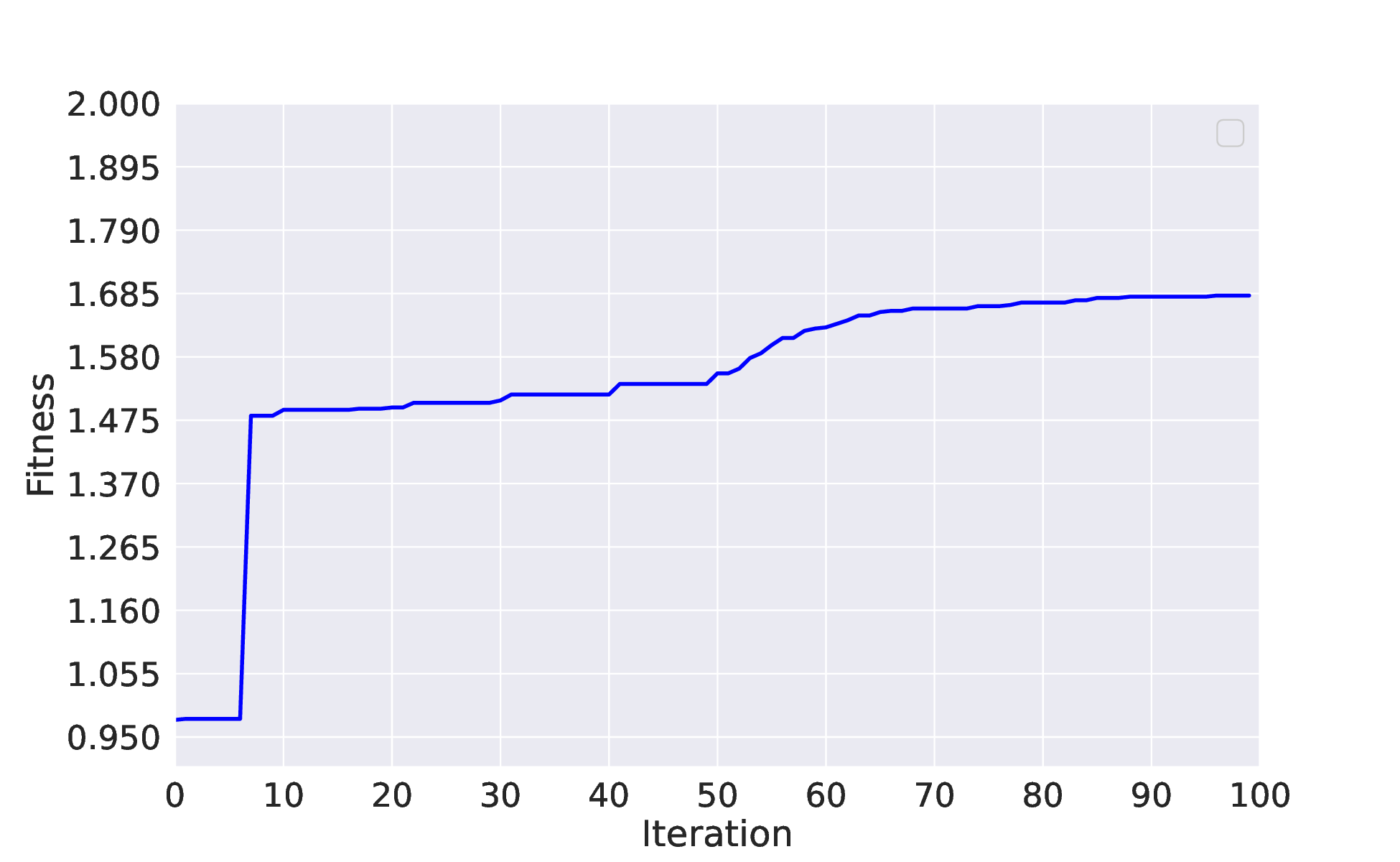}
	\caption{Fitness curve. The proposed approach successfully achieved an accuracy threshold of $0.98$ by the 8th iteration, after which it prioritized reducing the number of features by equation \ref{e8}.}
	\label{fig4}
\end{figure}

\begin{figure}
	\centering
	\includegraphics[width=0.5\textwidth]{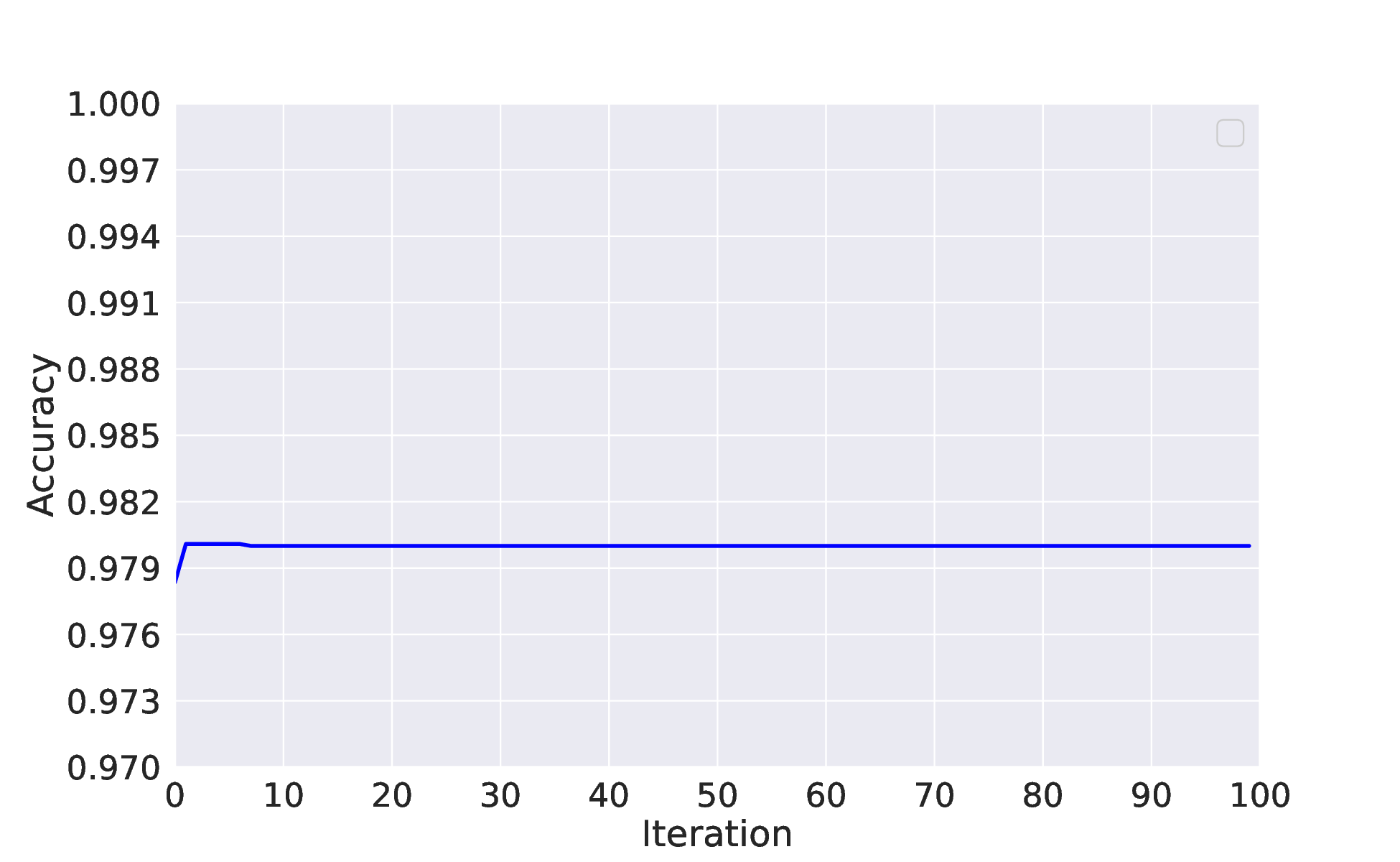}
	\caption{By the first iterations, the proposed approach effectively reached an accuracy threshold of 0.98. Subsequently, it aimed to maximize the fitness function defined in Equation \ref{e8} while maintaining the achieved accuracy level.}
	\label{fig7}
\end{figure}

The XOR PSO algorithm employed a population size of $100$ particles and conducted $100$ iterations to search for the optimal solution. Through this iterative process, the algorithm successfully identified the best solution consisting of $163$ features. Subsequently, a k-NN classifier is trained using these selected features, resulting in an impressive accuracy of $0.98$. The visual representation of these results can be observed in Figure \ref{fig3}. Figure \ref{fig4} displays the fitness curve, illustrating the changes in fitness values over the course of the optimization process. Table \ref{tab3} presents the performance metrics of the proposed method, including precision, recall, F1 score, and accuracy. Figure \ref{fig7} illustrates the achieved accuracy of the proposed method.
Table \ref{tab2} displays the comparison of execution times between the AAPSO method and the proposed method. The results indicate that each iteration of the AAPSO method requires approximately $28$ seconds, whereas the proposed PSO method completes an iteration in approximately $14$ seconds.

\begin{figure}
	\centering
	\includegraphics[width=0.5\textwidth]{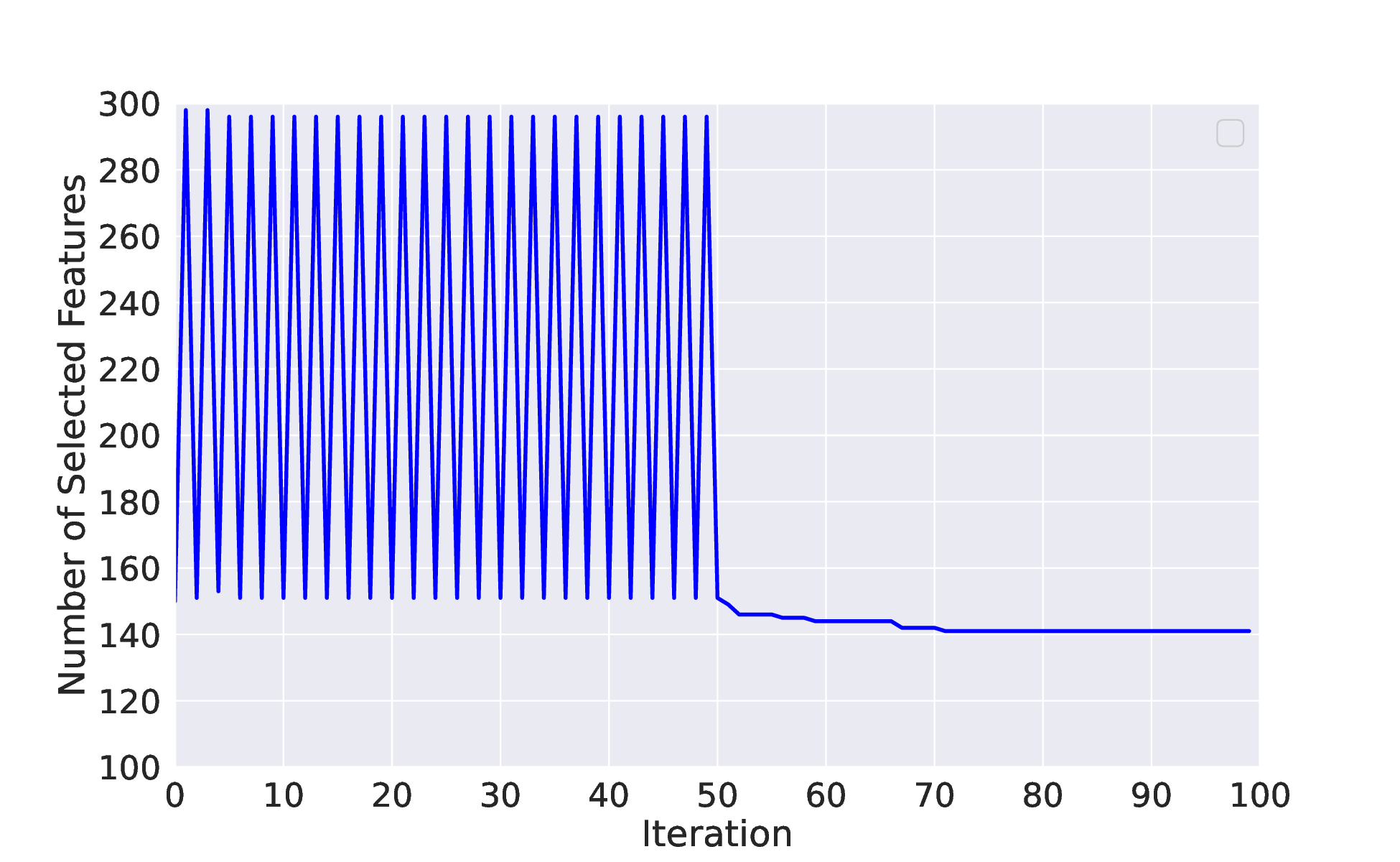}
	\caption{Visualizing the evolution of solution size for feature selection on ResNet50. The algorithm focused on exploration in first iterations by achieving the better accuracy. After achieving accuracy more than threshold, focused on decreasing the number of features (by equation \ref{e8}).}
	\label{fig8}
\end{figure}

\begin{figure}
	\centering
	\includegraphics[width=0.5\textwidth]{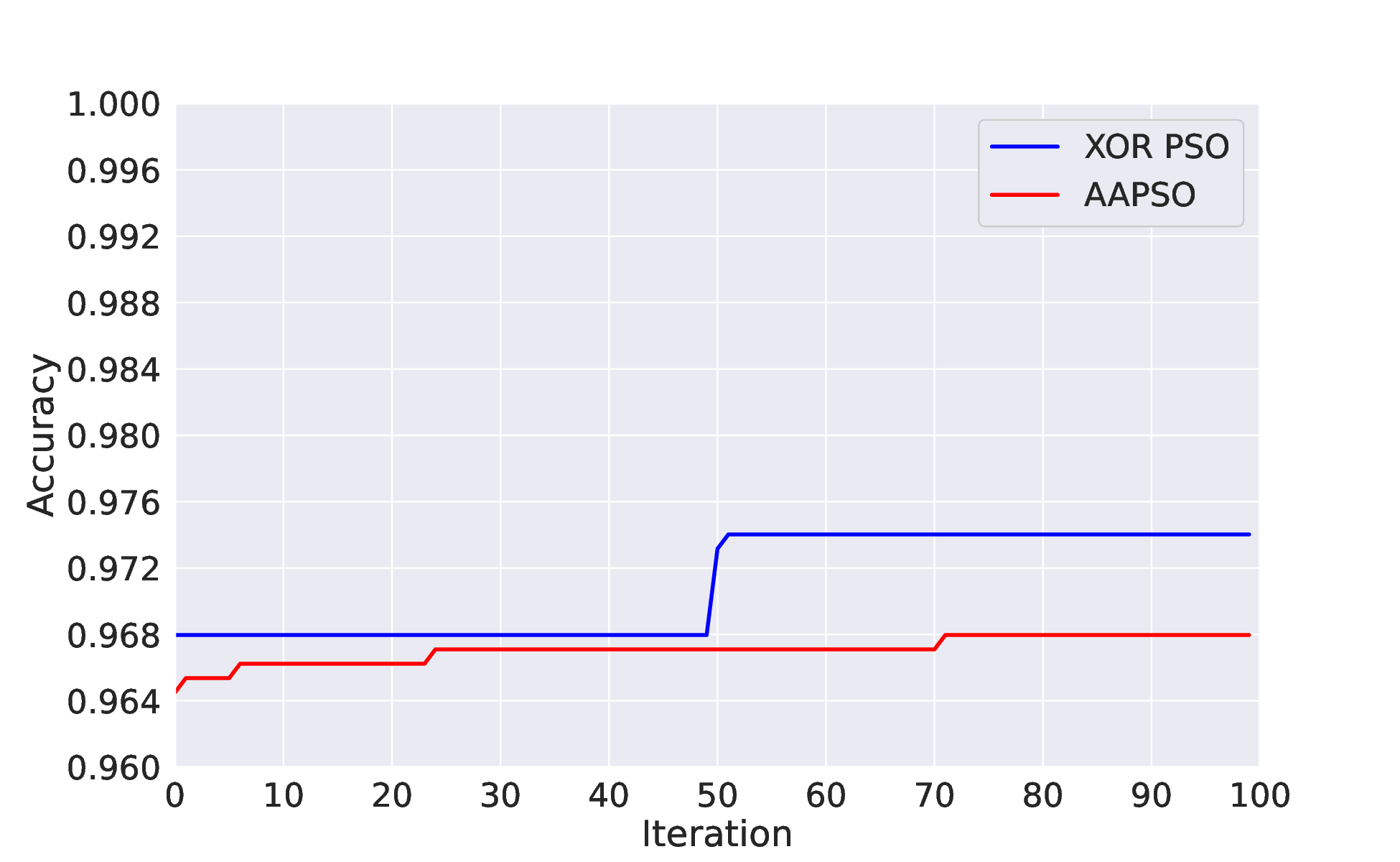}
	\caption{Comparative Analysis of Two Approaches on ResNet50 Extracted Features.}
	\label{fig9}
\end{figure}

In another experiment, ResNet50 were utilized as the CNN model and both the AAPSO and XOR PSO approaches performed on the extracted features. The XOR PSO method achieved an impressive accuracy of $97.4\%$ using only 140 features, whereas the AAPSO method achieved an accuracy of $96.8\%$ but required 224 features. The evolution of the best solution size in each iteration is depicted in Figure \ref{fig8}. As shown in Figure \ref{fig9}, XOR PSO could exceed the accuracy of $98\%$ in iteration $50$ and then focused on decreasing the number of features while AAPSO couldn't improve its accuracy as much.

\begin{table}[]
	\centering
	\caption{Results of proposed PSO method.}
	\label{tab3}
	\begin{tabular}{@{}lllll@{}}
		\toprule
		\textbf{Class} & \textbf{Precision} & \textbf{Recall} & \textbf{F1} & \textbf{Accuracy} \\ \midrule
		\textbf{0}     & 0.9691             & 0.9632          & 0.9662      &                   \\
		\textbf{1}     & 0.9856             & 0.9879          & 0.9867      &                   \\
		&                    &                 &             & 0.9810            \\ \bottomrule
	\end{tabular}
	
\end{table}

\begin{table}[ht]
	\centering
	\caption{Comparison of Time Between AAPSO and Proposed PSO}
	\label{tab2}
	\begin{tabular}{@{}lllll@{}}
		\toprule
		\multirow{2}{*}{\textbf{Iteration}} & \multicolumn{2}{c}{\textbf{Time (in seconds)}} \\ 
		& \textbf{AAPSO} & \textbf{XOR PSO} \\ \midrule
		\textbf{1} &27.55 & 14.95 \\ 
		\textbf{2} &28.54 & 14.44 \\ 
		\textbf{3} &29.55 & 14.59 \\ 
		\textbf{50} &29.32 & 14.52\\  
		\textbf{51} &28.97 & 14.43\\  
		\textbf{52} & 29.14& 14.53\\  
		\textbf{98} & 28.91& 13.39\\  
		\textbf{99} & 28.77& 12.98\\  
		\textbf{100} & 29.01& 12.79\\ 
		\textbf{AVG} & 27.92 & 13.88 \\ 
		\textbf{SUM} &2792.33 &1388.72 \\ \bottomrule
		
	\end{tabular}
\end{table}

\section{Conclusion}
Pneumonia remains a significant challenge in developing countries, where limited access to medical equipment and experts hinders effective detection and treatment. The AI-based algorithms holds great promise in addressing this issue. In this research, a RegNet model which is fine-tuned using a dataset of CXR images of children is selected. By incorporating the XOR PSO algorithm, improved accuracy in pneumonia detection was achieved. The algorithm effectively extracted $163$ features from the second last layer of the RegNet model, resulting in a significant increase in accuracy. Proposed approach achieved an impressive accuracy rate of $0.98$, indicating its potential for reliable pneumonia detection in CXR images. Furthermore, the XOR PSO algorithm demonstrated notable advantages over other PSO-based algorithms, particularly in terms of execution time and accuracy. The algorithm is simple and efficient, as it requires only one hyperparameter for initialization. This simplicity enhances its practicality and usability. By striking a balance between exploration and exploitation, this method achieved convergence on a highly accurate solution.

This research shows the effectiveness of the RegNet model fine-tuned with the XOR PSO algorithm for pneumonia detection in CXR images. This approach not only demonstrates enhanced accuracy but also exhibits improved efficiency compared to other PSO-based algorithms. These findings highlight the potential of AI-based algorithms in supporting healthcare initiatives by facilitating reliable and timely diagnosis of pneumonia.

\bibliographystyle{IEEEtran}
\bibliography{main_refs_new}

\begin{thebibliography}{10}
\providecommand{\url}[1]{#1}
\csname url@samestyle\endcsname
\providecommand{\newblock}{\relax}
\providecommand{\bibinfo}[2]{#2}
\providecommand{\BIBentrySTDinterwordspacing}{\spaceskip=0pt\relax}
\providecommand{\BIBentryALTinterwordstretchfactor}{4}
\providecommand{\BIBentryALTinterwordspacing}{\spaceskip=\fontdimen2\font plus
\BIBentryALTinterwordstretchfactor\fontdimen3\font minus
  \fontdimen4\font\relax}
\providecommand{\BIBforeignlanguage}[2]{{%
\expandafter\ifx\csname l@#1\endcsname\relax
\typeout{** WARNING: IEEEtran.bst: No hyphenation pattern has been}%
\typeout{** loaded for the language `#1'. Using the pattern for}%
\typeout{** the default language instead.}%
\else
\language=\csname l@#1\endcsname
\fi
#2}}
\providecommand{\BIBdecl}{\relax}
\BIBdecl

\bibitem{makhnevich2019clinical}
A.~Makhnevich, L.~Sinvani, S.~L. Cohen, K.~H. Feldhamer, M.~Zhang, M.~L.
  Lesser, and T.~G. McGinn, ``The clinical utility of chest radiography for
  identifying pneumonia: accounting for diagnostic uncertainty in radiology
  reports,'' \emph{American Journal of Roentgenology}, vol. 213, no.~6, pp.
  1207--1212, 2019.

\bibitem{Torres2021}
\BIBentryALTinterwordspacing
A.~Torres, C.~Cilloniz, M.~S. Niederman, R.~Menéndez, J.~D. Chalmers, R.~G.
  Wunderink, and T.~van~der Poll, ``Pneumonia,'' \emph{Nature Reviews Disease
  Primers}, vol.~7, no.~1, p.~25, 2021. [Online]. Available:
  \url{https://doi.org/10.1038/s41572-021-00259-0}
\BIBentrySTDinterwordspacing

\bibitem{wang2021deep}
G.~Wang, X.~Liu, J.~Shen, C.~Wang, Z.~Li, L.~Ye, X.~Wu, T.~Chen, K.~Wang,
  X.~Zhang \emph{et~al.}, ``A deep-learning pipeline for the diagnosis and
  discrimination of viral, non-viral and covid-19 pneumonia from chest x-ray
  images,'' \emph{Nature biomedical engineering}, vol.~5, no.~6, pp. 509--521,
  2021.

\bibitem{huang2022lightweight}
M.-L. Huang and Y.-C. Liao, ``A lightweight cnn-based network on covid-19
  detection using x-ray and ct images,'' \emph{Computers in Biology and
  Medicine}, vol. 146, p. 105604, 2022.

\bibitem{gayathri2022computer}
J.~Gayathri, B.~Abraham, M.~Sujarani, and M.~S. Nair, ``A computer-aided
  diagnosis system for the classification of covid-19 and non-covid-19
  pneumonia on chest x-ray images by integrating cnn with sparse autoencoder
  and feed forward neural network,'' \emph{Computers in biology and medicine},
  vol. 141, p. 105134, 2022.

\bibitem{pramanik2022adaptive}
R.~Pramanik, S.~Sarkar, and R.~Sarkar, ``An adaptive and altruistic pso-based
  deep feature selection method for pneumonia detection from chest x-rays,''
  \emph{Applied Soft Computing}, vol. 128, p. 109464, 2022.

\bibitem{narin2021accurate}
A.~Narin, ``Accurate detection of covid-19 using deep features based on x-ray
  images and feature selection methods,'' \emph{Computers in Biology and
  Medicine}, vol. 137, p. 104771, 2021.

\bibitem{canayaz2021mh}
M.~Canayaz, ``Mh-covidnet: Diagnosis of covid-19 using deep neural networks and
  meta-heuristic-based feature selection on x-ray images,'' \emph{Biomedical
  Signal Processing and Control}, vol.~64, p. 102257, 2021.

\bibitem{xu2022regnet}
J.~Xu, Y.~Pan, X.~Pan, S.~Hoi, Z.~Yi, and Z.~Xu, ``Regnet: self-regulated
  network for image classification,'' \emph{IEEE Transactions on Neural
  Networks and Learning Systems}, 2022.

\bibitem{kermany2018identifying}
D.~S. Kermany, M.~Goldbaum, W.~Cai, C.~C. Valentim, H.~Liang, S.~L. Baxter,
  A.~McKeown, G.~Yang, X.~Wu, F.~Yan \emph{et~al.}, ``Identifying medical
  diagnoses and treatable diseases by image-based deep learning,'' \emph{cell},
  vol. 172, no.~5, pp. 1122--1131, 2018.

\bibitem{Szegedy_Ioffe_Vanhoucke_Alemi_2017}
\BIBentryALTinterwordspacing
C.~Szegedy, S.~Ioffe, V.~Vanhoucke, and A.~Alemi, ``Inception-v4,
  inception-resnet and the impact of residual connections on learning,''
  \emph{Proceedings of the AAAI Conference on Artificial Intelligence},
  vol.~31, no.~1, Feb. 2017. [Online]. Available:
  \url{https://ojs.aaai.org/index.php/AAAI/article/view/11231}
\BIBentrySTDinterwordspacing

\bibitem{Xie2016}
S.~Xie, R.~Girshick, P.~Dollár, Z.~Tu, and K.~He, ``Aggregated residual
  transformations for deep neural networks,'' \emph{arXiv preprint
  arXiv:1611.05431}, 2016.

\bibitem{kennedy1995particle}
J.~Kennedy and R.~Eberhart, ``Particle swarm optimization,'' in
  \emph{Proceedings of ICNN'95-international conference on neural networks},
  vol.~4.\hskip 1em plus 0.5em minus 0.4em\relax IEEE, 1995, pp. 1942--1948.

\bibitem{he2016deep}
K.~He, X.~Zhang, S.~Ren, and J.~Sun, ``Deep residual learning for image
  recognition,'' in \emph{Proceedings of the IEEE conference on computer vision
  and pattern recognition}, 2016, pp. 770--778.

\bibitem{tan2019efficientnet}
M.~Tan and Q.~Le, ``Efficientnet: Rethinking model scaling for convolutional
  neural networks,'' in \emph{International conference on machine
  learning}.\hskip 1em plus 0.5em minus 0.4em\relax PMLR, 2019, pp. 6105--6114.

\end{thebibliography}

\end{document}